\documentclass[%
reprint,
amsmath,amssymb,
aps,
]{revtex4-2}
\usepackage{graphicx}
\usepackage[colorlinks=true,linkcolor=blue,anchorcolor=blue, citecolor=blue,urlcolor=blue]{hyperref}

\begin{document}
	\preprint{APS/123-QED}

\title{Higher order topological state induced by $d$-wave competing orders in high-T$_c$ superconductor based heterostructure}

\author{Xiaoming Wang$^{1}$}
\thanks{These two authors contributed equally to this work.}
\author{Yu-Xuan Li$^2$}
\thanks{These two authors contributed equally to this work.}
\author{Tao Zhou$^{1}$}
\email{Corresponding author: tzhou@scnu.edu.cn}
\affiliation{$^1$Guangdong Provincial Key Laboratory of Quantum Engineering and Quantum Materials, School of Physics and Telecommunication Engineering, Guangdong-Hong Kong Joint Laboratory of Quantum Matter, and Frontier Research Institute for Physics,
South China Normal University, Guangzhou 510006, China\\
$^2$Centre for Quantum Physics, Key Laboratory of Advanced Optoelectronic Quantum Architecture and Measurement (MOE), School of Physics, Beijing Institute of Technology, Beijing 100081, China
}

\begin{abstract}
	
We introduce a two-dimensional Chern insulator in proximity to a $d$-wave pseudogap state of the high-T$_c$ superconducting material as an effective platform to realize the higher order topological system. The
proximity-induced $d$-density-wave (DDW) order in the Chern insulator layer serves as an effective mass. The edge states will be fully gapped by this DDW order. In the real space, the sign of the DDW order parameter
changes at the system corners due to the $d$-wave factor, leading to the gapless corner states, indicating that this system may be in a higher order topological state. The higher order topology in this coupled system is confirmed based on the calculation of the edge polarization and the quadrupole moment. In the superconducting state where the superconducting order and the DDW order coexist, the Majorana corner states emerge.
\end{abstract}

\maketitle

\section{Introduction}

Higher order topology has gained much attention in recent years~\cite{re32,re9,re19,PhysRevB.98.201114,PhysRevB.103.L201115,re117,re122,re130,PhysRevB.100.245134,PhysRevResearch.1.032048,PhysRevB.100.245135,PhysRevLett.125.166801,re190,re21,PhysRevB.99.041301,Park_2019,Sheng_2019,PhysRevLett.122.256402,Noguchi_2021,RN4,re34,re100,re35,re37,PhysRevB.103.024517,Xue_2018,Serra_Garcia_2018,Ni_2018,Peterson_2018}. For a $d$-dimensional n-th order topological material, the system is fully gapped at its $d-1$ dimensional boundaries, while it hosts gapless states at its d-n dimensional boundaries $(n\geq 2)$. In a two-dimensional higher order topological material, the gapless modes emerge at the corners between different fully gapped one-dimensional edges. In the past several years, the higher order topology has been proposed to exist in the three-dimensional Bismuth material~\cite{re19,Noguchi_2021}, the twisted angle graphene~\cite{Park_2019}, the graphdiyne~\cite{Sheng_2019}, and some heterostructure systems and artificial systems~\cite{RN4,re34,re100,re35,re37,PhysRevB.103.024517,Xue_2018,Serra_Garcia_2018,Ni_2018,Peterson_2018}.

A number of proposals for the realization of higher
order topological superconductors demand unconventional superconducting pairing term, for instance, the $d$-wave pairing term~\cite{RN4,re34,re100,re35,re37,PhysRevB.103.024517}.
For a heterostructure with an additional superconducting pairing term being added to a first order topological system, an energy gap may be induced by the superconducting pairing term. Then, the original gapless system edges may be fully gapped.
For a $d$-wave superconductor, the sign of the pairing term changes at the system corners. As a result, the effective mass may change and the band inversion occurs at these corners, leading to the possible Majorana corner states.
Now it is well accepted that the cuprate high-T$_c$ superconducting material has the $d$-wave pairing symmetry. Therefore, the heterostructure with the high-T$_c$ superconducting material may provide a useful platform to realize the higher order topological superconductor or the Majorana corner states.

The high-T$_c$ superconductivity in the family of the cuprate materials has been studied for more than three decades.
One of the most important issues is the pseudogap behaviour in the underdoped region of the phase diagram~\cite{Mueller_2017}.
So far the origin of the pseudogap is still an open question. A possible explanation is that the pseudogap may be due to a certain competing order~\cite{PhysRevB.62.4880,PhysRevB.63.094503,PhysRevB.74.184515,RevModPhys.87.457}. In the topological superconducting system with the high-T$_c$ superconductor platform, the interplay between the competing order and the topological behaviour is also of interest~\cite{PhysRevB.99.104517,Li2022}. The additional energy gap induced by the competing order may change the topological behaviour, so that the topological behaviour may be used to detect the competing orders~\cite{PhysRevB.99.104517}. And recently it was indicated that a charge density wave order may be used to modulate the gap size and the Majorana zero modes~\cite{Li2022}.

Experimentally, it was indicated that the pseudogap in the high-T$_c$ superconducting material has the similar momentum dependence with the superconducting gap. Therefore, generally a $d$-wave type factor may exist in the possible competing pseudogap order~\cite{PhysRevB.62.4880,PhysRevB.63.094503,PhysRevB.74.184515,PNAS,Comin2015,Forgan2015}.
The existence of the $d$-wave factor may lead to the mass change and the band inversion at the system corners. Therefore, it is naturally to ask an interesting question ``can the $d$-wave type pseudogap in high-T$_c$ superconducting materials be used to realize the higher order topology?''. If the answer is positive, more effective proposals may be brought forward to realize the higher order topological states. On the other hand, this may be used to detect and resolve different competing orders of the high-T$_c$ superconductors, providing more insight for the physical origin of the
pseudogap behaviour in the high-T$_c$ superconductors.

In this paper, we consider a heterostructure of a two-dimensional Chern insulator in proximity of a high-T$_c$ superconducting material with the $d$-density wave (DDW) order~\cite{supp}.
Our numerical results indicate that the gapless edge states of a Chern insulator are gapped out by the DDW order. The gapless states emerge at the system corners. These results propose that this coupled system is indeed a higher order topological system. The higher order topology is confirmed further through calculating the edge polarization with the Wilson loop method and the quadrupole moment.

The rest of the paper is organized as follows. In Sec. II,
we introduce the model and present the relevant formalism.
In Sec. III, we report numerical calculations and discuss the
obtained results. Finally, we give a brief summary in Sec. IV.

\section{Model and formalism}
We start with a Hamiltonian in a two-dimensional square lattice considering the coupling of a Chern insulator and an additional DDW order term~\cite{PhysRevB.62.4880,PhysRevB.63.094503}, with the whole Hamiltonian being expressed as,
\begin{equation}
	H=H_C+H_{D}.
\end{equation}

$H_C$ is a two-band model describing the Chern insulator~\cite{PhysRevB.74.045125,PhysRevB.78.195424},
\begin{equation}
	H_C=\sum_{{\bf k}\sigma}\sigma \varepsilon_{\bf k}  c^\dagger_{{\bf k}\sigma}c_{{\bf k}\sigma}+\sum_{\bf k}(\lambda_{\bf k}c^\dagger_{{\bf k}\uparrow}c_{{\bf k}\downarrow}+h.c.),
\end{equation}
with $\sigma$ takes $+$ and $-$ for the spin-up and spin-down quasiparticles, respectively.
 $\varepsilon_{\bf k}=-2 t(\cos k_x+\cos k_y)-m$ and $\lambda_{\bf k}=2\lambda_0(\sin k_x+i\sin k_y)$, represent the spin polarized hopping and the spin-orbital coupling term, respectively.
$m$ is the effective Zeeman field strength. The above Chern insulator model may be realized experimentally in the Hg$_{1-x}$Mn$_x$Te/Cd$_{1-x}$Mn$_x$Te quantum well~\cite{PhysRevB.74.045125} or the magnetic topological insulators (Bi,Sb)$_2$Te$_3$~\cite{sci1234414} and MnBi$_{8}$Te$_{13}$~\cite{Hu_2020}.

$H_{D}$ is a proximity induced DDW term, with
\begin{eqnarray}
	H_{D}=\sum_{\bf k}(\Delta_{\bf k} c^\dagger_{{\bf k}\sigma}c_{{\bf k}+{\bf Q}\sigma}+h.c.),
\end{eqnarray}
Following Refs.~\cite{PhysRevB.62.4880,PhysRevB.63.094503}, we consider the $d$-wave form factor with $\Delta_{\bf k}=2i\Delta_0(\cos k_x-\cos k_y)$ and the wavevector ${\bf Q}$ is taken as $(\pi,\pi)$.

 To study the edge states of the system, we would like to consider a cylinder geometry with
 considering
 the periodic boundary condition along the $y$ direction and the open boundary condition along the $x$-direction.
 Then we define a partial
 Fourier transformation along the $x$-direction with $c^\dagger_{{\bf k}\sigma}=\frac{1}{{\sqrt {N_x}}}\sum_x c^\dagger_{{k_y}\sigma}(x)e^{ik_x x}$. The Hamiltonian is rewritten as,
 \begin{eqnarray}
 H_C=&-t\sum\limits_{k_y,x,\sigma}[\sigma c^\dagger_{k_y\sigma}(x)c_{k_y\sigma}(x+1)+h.c.]\nonumber\\&-\lambda_0\sum\limits_{k_y,x}[ic^\dagger_{k_y\uparrow}(x)c_{k_y\downarrow}(x+1)+h.c.]
 \nonumber\\&+\lambda_0\sum\limits_{k_y,x}[ic^\dagger_{k_y\uparrow}(x)c_{k_y\downarrow}(x-1)+h.c.]
 \nonumber\\&-\sum\limits_{k_y,x,\sigma}
 \sigma(m+2t\cos k_y)c^\dagger_{k_y\sigma}(x)c_{k_y\sigma}(x)\nonumber\\&+2\lambda_0\sum\limits_{k_y,x}[i\sin k_yc^\dagger_{k_y\uparrow}(x)c_{k_y\downarrow}(x)+h.c.],
\end{eqnarray}
and
 \begin{eqnarray}
 H_{D}=&\sum\limits_{k_yx\sigma}[(-1)^{R_x}i\Delta_{0} c^\dagger_{{k_y\sigma}}(x)c_{{k_y+\pi\sigma}}(x+1)+h.c.]\nonumber\\&-2\sum\limits_{k_yx\sigma}[(-1)^{R_x}i\Delta_{0}\cos k_y  c^\dagger_{{k_y\sigma}}(x)c_{{k_y+\pi\sigma}}(x)+h.c.].\nonumber\\
 \end{eqnarray}

 The above Hamiltonian can be rewritten as a $4N_x\times 4N_x$ matrix $\hat{M}({k_y})$ with $H=\sum_{k_y}\Psi^{\dagger}(k_y)\hat{M}(k_y)\Psi(k_y)$. $N_{x}$ is the number of the sites along the $x$ direction.
 The vector $\Psi^\dagger(k_{y})$ is expressed as,
 \begin{equation}
 		\Psi^\dagger(k_{y}) =(C_1,C_2, \cdots,C_{N_x}),
 \end{equation}
 with $C_i=(c^\dagger_{i\uparrow}(k_y),c^\dagger_{i\downarrow}(k_y),c^\dagger_{i\uparrow}(k_y+\pi),c^\dagger_{i\downarrow}(k_y+\pi))$.

The higher order topology of the system can be studied through calculating the edge polarization~\cite{re32,re9,re19,PhysRevB.98.201114,PhysRevB.103.L201115}. We define a Wilson loop matrix with the elements being written as,
\begin{eqnarray}
[W(x,k_{y})]_{ij}=&&\langle u^{i}(x,k_{y}) | P^{1}(x,k_{y}+\Delta k_{y}) \nonumber \\
&&  P^{2}(x,k_{y}+2\Delta k_{y}) \nonumber \\
&&\dots  P^{N-2}(x,k_{y}+(N-2)\Delta k_{y}) \nonumber \\
&&| u^{j}(x,k_{y})\rangle.
\end{eqnarray}
$P^{l}(x,k_y+l\Delta k_y)$ ($l=1,2,3,\dots,N-2$) is the projection operator with $P^{l}(x,k_y+l\Delta k_y)=\sum_n \left| u^n({x,k_y+l\Delta k_y})\right\rangle \left\langle  u^n(x,k_y+l\Delta k_y)\right| $. $\left| u^n(x,k_y+l\Delta k_y)\right\rangle$ are the eigenvectors of the matrix $\hat{M}({k_y})$, with only the occupied bands being considered. Diagonalizing the Wilson loop matrix, we have,
\begin{eqnarray}
W(x,k_{y})\left| \nu^{j}_{k_{y}}\right\rangle =e^{i2\pi\nu^{j}_{x}}\left| \nu^{j}_{k_{y}}\right\rangle,
\end{eqnarray}
where $j$ is corresponding to the occupied band index. $\nu^{j}_{x}$ is defined as a Wannier spectrum.
The tangential polarization as a function of lattice site $R_{x}$ is given as,
\begin{eqnarray}
p_{y}(R_{x})=\sum^{occu} \rho^{j}(R_{x})\nu^{j}_{x},
\end{eqnarray}
where $\rho^{j}(R_{x})=\frac{1}{N_{x}}\sum_{k_{y},\sigma,n}\left| [u^{j}(R_{x},k_{y})]^{n,\sigma}[\nu_{k_{y}}^{j}]^{n} \right|^{2} $. The edge polarization at the $x$ direction is given as
\begin{equation}
p^{edge,x}_{y}=\sum_{R_{x}=1}^{R_{x}=N_{x}/2}	p_{y}(R_{x}).
\end{equation}

 For a higher order topological system, the gapless states are expected to emerge at the system corners.
 In this case one needs to consider the open boundary condition along both two directions. The Hamiltonian should be expressed in the real space through
  a full Fourier transformation, with
  \begin{eqnarray}
 H=&-t\sum\limits_{{\bf i},{\alpha},\sigma}[\sigma c^\dagger_{{\bf i}\sigma}c_{{\bf i}+\hat{\alpha}\sigma}+h.c.]-\sum\limits_{{\bf i},\sigma}\sigma m c^\dagger_{{\bf i}\sigma}c_{{\bf i}\sigma}\nonumber\\&-\lambda_0\sum\limits_{\bf i}[ic^\dagger_{{\bf i}\uparrow}c_{{\bf i}+\hat{x}\downarrow}-ic^\dagger_{{\bf i}\uparrow}c_{{\bf i}-\hat{x}\downarrow}+h.c.]\nonumber\\
&+\lambda_0\sum\limits_{\bf i}[c^\dagger_{{\bf i}\uparrow}c_{{\bf i}+\hat{y}\downarrow}-c^\dagger_{{\bf i}\uparrow}c_{{\bf i}-\hat{y}\downarrow}+h.c.] \nonumber\\&-\sum\limits_{{\bf i}\sigma}[(-1)^{R_x+R_y}i\Delta_{0} (c^\dagger_{{\bf i}\sigma}c_{{\bf i}+\hat{x}\sigma}-c^\dagger_{{\bf i}\sigma}c_{{\bf i}+\hat{y}\sigma})\nonumber\\&+h.c.].
 \end{eqnarray}

The above real space Hamiltonian can be written as a $2N\times 2N$ matrix with $H=\Psi^{\dagger}\hat{M}\Psi$ ($N=N_xN_y$ is the number of the sites). Then following Refs.~\cite{PhysRevB.100.245134,PhysRevResearch.1.032048,PhysRevB.100.245135,PhysRevLett.125.166801}, we define the quadrupole moment, with

\begin{eqnarray}
q_{x y}=\frac{1}{2 \pi} \operatorname{Im} \log \left[\operatorname{det}\left(U^{\dagger} \hat{Q} U\right) \sqrt{\operatorname{det}\left(\hat{Q}^{\dagger}\right)}\right],
\end{eqnarray}
where $\hat{Q} \equiv \exp \left[i 2 \pi \hat{q}_{x y}\right]$. $\hat{q}_{x y}$ is the position operator and can be written as a diagonalized matrix with the elements being $R_x R_y/N$. The matrix $U$ is constructed by column-wise packing all of occupied eigenstates.

The Green's function can be defined through diagonalizing the above Hamiltonian matrix $\hat{M}$. The matrix elements of the Green's function is expressed as,

\begin{eqnarray}
	G(\omega)_{ij}=
	\sum_{n}\dfrac{u_{in} u^{\dagger}_{jn}}{\omega-E_{n}+i\Gamma},
\end{eqnarray}
where $u_{in}$ and $E_{n}$ are the eigenvectors and the eigenvalues of the Hamiltonian matrix.

In the real space, the local density of states at the site $i$ is calculated from the real space Green's function, with
\begin{eqnarray}
	\rho_{i}(\omega)=-\frac{1}{\pi}\sum^2_{p=1}\mathrm{Im}G(\omega)_{l+p,l+p},
\end{eqnarray}
where $l=2(i-1)$.

 In the following presented results, the parameters are set as $t=1$, $\lambda_0=1$, $m=1$, and $\Delta_{0}=0.4$. We have checked numerically that our main conclusions are not sensitive with these parameters.

\section{results and discussion}

\begin{figure}
	\includegraphics[width=8.5cm,height=8.5cm]{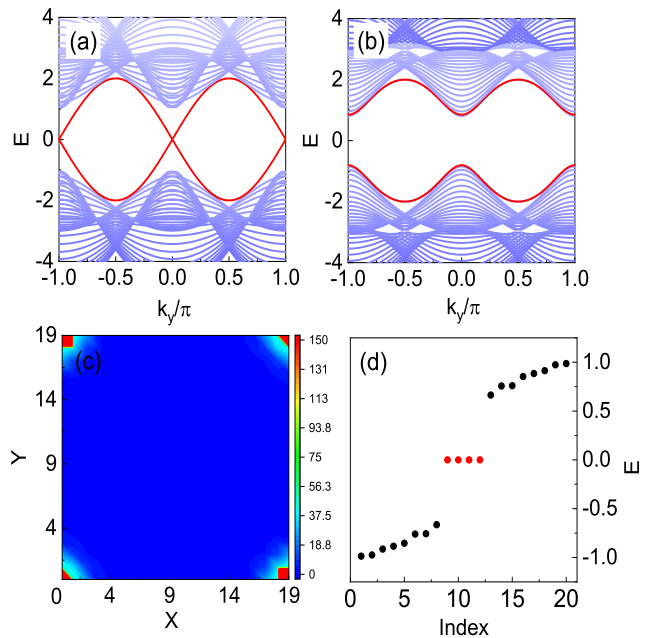}
	\caption{(a) The energy bands considering the open boundary condition along the $x$-direction with $\Delta_0=0$. (b) The energy bands in the presence of the DDW modulation. (c) The zero energy LDOS in the presence of the DDW order in a $20\times 20$ lattice with open boundaries along both the $x$ and the $y$ directions. (d) The corresponding eigenvalues of the Hamiltonian in the $20\times 20$ lattice.
	}
	\label{fig:1}
\end{figure}

We first study the possible edge states with considering the open boundary condition along the $x$ direction and periodic boundary condition along the $y$ direction. The energy bands obtained by diagonalizing the $4N_x\times4N_x$ Hamiltonian matrix $\hat{M}(k_y)$
with $\Delta_0=0$ and $\Delta_0=0.4$ are presented in Figs.~1(a) and 1(b), respectively. Without the DDW term ($\Delta_0=0$), the matrix $\hat{M}(k_y)$ can be block diagonalized as two $2N_x\times2N_x$ matrices, expressed as $H(k_y)$ and $H(k_y+\pi)$, respectively. With the cylinder geometry, the energy bands are gapless and the chiral edge states exist, as is seen in Fig.~1(a). When the DDW term is added to the system with $\Delta_0=0.4$, as is seen in Figs.~1(b), it is clearly that the chiral edge states are fully gapped. This indicates that in the presence of the DDW term, the whole system is not a first order topological system.

We now consider the open boundary condition along both $x$ and $y$ directions and diagonalizing the $2N\times 2N$ Matrix $\hat{M}$. The zero energy LDOS spectrum is displayed in Fig.~1(c). As is seen, the zero energy corner states indeed exist. We present the the eigenvalues of the Matrix in Fig.~1(d). Four zero energy eigenvalues protected by an energy gap are seen clearly. The numerical results indicate that this system may indeed be a higher order topological system.

An intuitive picture for the higher order topology can be provided based on the edge theory~\cite{RN4}. When the DDW order is zero, the linear gapless states emerge at the system edge as the effective Zeeman field $0<\mid m \mid <2$~\cite{PhysRevB.78.195424}. As $m=0$, the system is a two-dimensional Weyl semimetal and the linear gapless states also exist. Therefore, without the DDW term, the system edges are gapless as $\mid m \mid <2$.
In the presence of the DDW term, as is shown in Fig.~1, with the cylinder geometry, the system edges are fully gapped and the DDW term acts as an effective mass term. Similar to the $d$-wave superconducting order discussed in Ref.~\cite{RN4}, here due to the existence of the $d$-wave factor in the DDW order, the sign of the DDW term changes at the system corners, leading to the gapless corner states, as is shown in Fig.~1(c).

\begin{figure}
	\includegraphics[width=8.5cm,height=4.5cm]{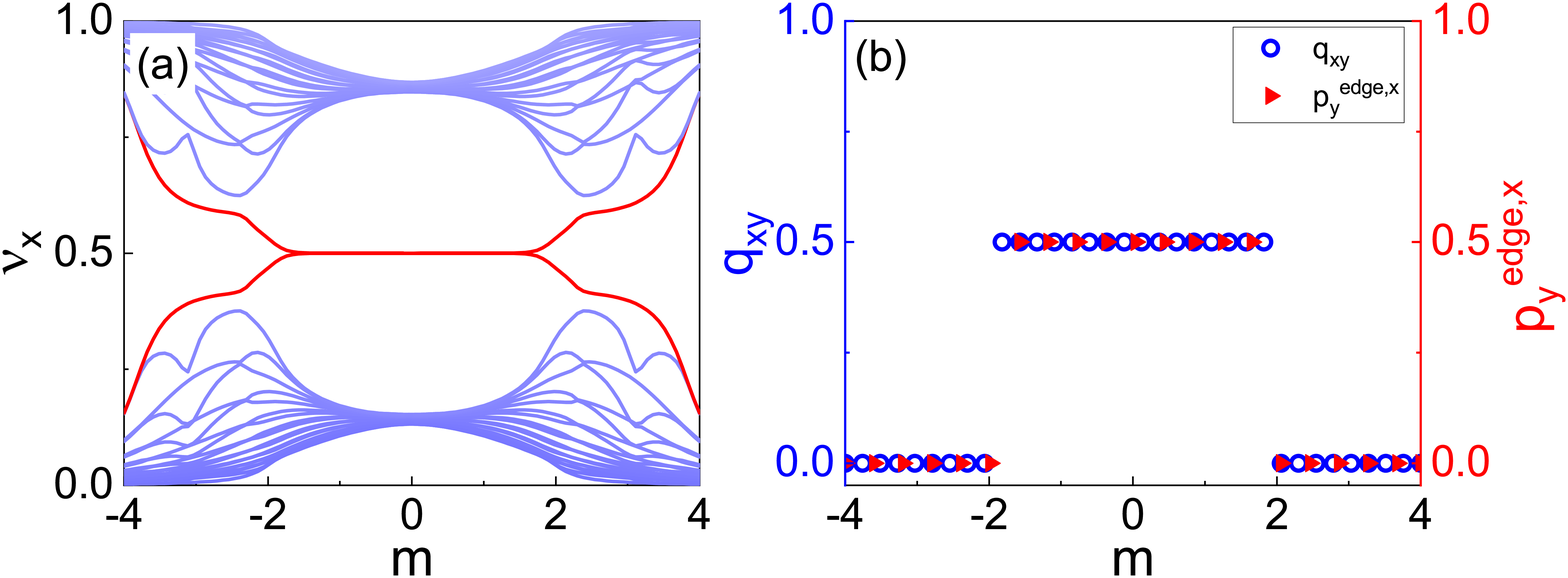}
	\caption{(a) The Wannier spectral $\nu_{x}$ as a function of the effective Zeeman field $m$. (b) The edge polarization ($p^{edge,x}_{y}$,$p^{edge,y}_{x}$) and quadrupole moments $q_{xy}$ as a function of $m$.
	}
	\label{fig:2}
\end{figure}

Now let us verify the high order topology of this system through calculating the edge polarization~\cite{re32,re9,re19,PhysRevB.98.201114,PhysRevB.103.L201115} and the quadrupole moment~\cite{PhysRevB.100.245134,PhysRevResearch.1.032048,PhysRevB.100.245135,PhysRevLett.125.166801}. With a cylinder geometry, the Wannier spectrum is obtained through diagonalizing the Wilson loop matrix [Eq.(8)]. The Wannier spectrum $\nu_x$ as a function of the effective Zeeman field $m$ is plotted in Fig.~2(a). The edge polarization calculated from Eq.(10) and the quadrupole moment from Eq.(12) are plotted in Fig.~2(b). As is seen, as $\mid m \mid<2$,  the Wannier spectra
is half quantized to 0.5, leading to the half quantized edge polarization at this parameter regime.
The numerical results of the quadrupole moment also obtain the same conclusion.
These results confirm that the system is
indeed a higher order topological one~\cite{re32,re9,re19,PhysRevB.98.201114,PhysRevB.103.L201115,PhysRevB.100.245134,PhysRevResearch.1.032048,PhysRevB.100.245135,PhysRevLett.125.166801}.

It was pointed out that there are some difficulties for the method of calculating the quadrupole moment from Eq.(12), namely, the improper choice of the origin point and using odd number lattice size may lead to quadruple moments being at zero or not quantized, even when the system is a second order topological one~\cite{PhysRevB.100.245133}. Here the origin point is not arbitrarily chosen. It is set at the site $(R_{x},R_{y})=(1,1)$. The system size is chosen as an even number along both the $x$ and $y$ directions. The difficulties proposed in Ref.~\cite{PhysRevB.100.245133} is partly avoided.
Moreover, it was proposed that the quadruple moments can be calculated through the corner charge and the edge polarizations~\cite{re32,re9}. We have checked numerically that with this method, the obtained quadrupole moment is the same with that obtained from Eq.(12)~\cite{supp}.

In a high-T$_c$ superconducting material, the superconducting order and the possible pseudogap order may coexist in the underdoped region~\cite{Mueller_2017}. It is useful to study the robustness of the corner states in presence of the superconducting order. Considering an additional $d$-wave superconducting pairing term with $H_{dSC}=\Delta_d \sum_{\bf k}[(\cos k_x-\cos k_y) c^\dagger_{{\bf k}\uparrow}c^\dagger_{-{\bf k}\downarrow}+h.c.]$. For the cylinder geometry with the open boundary along the $x$-direction, the whole Hamiltonian includes both the electron part and the hole part. It can be written as the $8N_x\times 8N_x$ matrix.
In the real space with $N$ lattice sites, the Hamiltonian can be written as a $4N\times4N$ matrix.
In presence of the $d$-wave superconducting order, we plotted the energy bands of the Hamiltonian with the cylinder geometry without and with the DDW order in Figs.~3(a) and 3(b), respectively.
The intensity plot of the zero energy LDOS with the coexistence of the superconducting order and the DDW order in the real space is presented in Fig.~3(c). The corresponding low energy eigenvalues are plotted in Fig.~3(d). Here the energy spectra are qualitatively the same when the superconducting order exists, namely,
in the pure superconducting state ($\Delta_d\neq 0$ and $\Delta_0=0$), the system has gapless edge states with a cylinder geometry. In the presence of the DDW term,
an energy gap is opened by the DDW order. In the real space with the open boundaries, the gapless corner states emerge. These results indicate that the system is still a second order topological system when the superconducting order and the DDW order coexist. And in this case, the gapless corner states are the Majorana zero modes.

\begin{figure}
	\includegraphics[width=8.5cm,height=8.5cm]{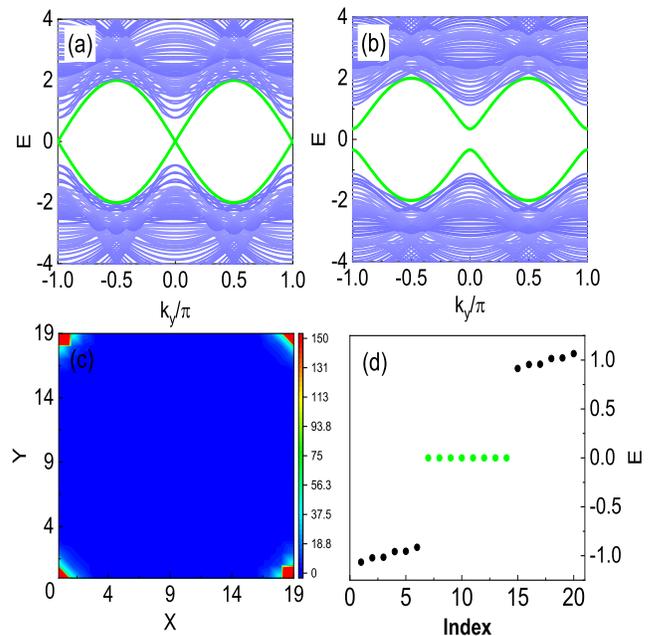}
	\caption{Similar to Fig.~1 but in presence of an additional $d$-wave pairing order with $\Delta_{d}=0.5$
	}
	\label{fig:3}
\end{figure}

Due to the particle-hole symmetry, there are eight zero energy eigenvalues for the real space Hamiltonian matrix, as is seen in Fig.~3(d), indicating that there are two zero energy modes at each corner. This is consistent with the Chern number of the system when the DDW order is absent.
For the Chern insulator system, the Chern number is 1 as $0<m<2$~\cite{PhysRevB.78.195424}. Then with the DDW
order, each corner has one gapless corner mode. With the $d$-wave superconducting order, the system becomes a topological superconductor~\cite{PhysRevB.82.184516,PhysRevB.92.064520,PhysRevLett.112.086401,PhysRevLett.120.017001}. Due to the particle-hole symmetry, the Chern number becomes 2. As a result, with the DDW
order, each corner contains two gapless Majorana zero modes. 
Note that these two zero modes are unstable. They are not protected by a certain symmetry. 
We have numerically checked that the zero modes disappear and the energy bands become fully gapped with an additional chemical potential term. Previously, it was discussed and verified that without the symmetry protection, the Majorana zero modes with the equal spin at the same site may exist, while they are indeed unstable. They may interfere with each other and annihilate as the finite energy quasiparticles~\cite{Wang_2022,doi:10.1126/sciadv.aaz8367,PhysRevB.100.205126,PhysRevLett.121.126402}.

On the other hand, previously, the superconductor/Chern-insulator coupled systems have been studied intensively~\cite{PhysRevB.82.184516,PhysRevB.92.064520,PhysRevLett.112.086401,PhysRevLett.120.017001}. The Chern number depends on the chemical potential and the magnitude of the superconducting order parameter.
A stable topological superconducting phase with the Chern number being 1 can indeed be realized with proper parameters. We have checked numerically that this stable topological superconductor would also become a higher order one in the presence of an additional DDW term. In addition, in this case, the open system has only one stable Majorana zero mode at each corner.

Finally, we would discuss the significance and outlook of our present work. Firstly, we here proposed and demonstrated that the higher order topology can indeed be realized through considering the $d$-wave form pseudogap order with the high-T$_c$ superconductor platform. This is of interest and worth further studies. Secondly, 
our present proposal can be extended to other topological systems or other platforms. For instance, we have checked that the higher order topological superconducting state can be realized for a topological-superconductor/DDW coupled system. On the other hand,
the DDW state is often called as the staggered flux state~\cite{PhysRevB.62.4880}, which can be easily realized and controlled in the cold atom system~\cite{PhysRevLett.107.255301}. 
Also, the pairing state, the spin-orbital coupling, and various topological states have been successfully realized 
in the cold atom system~\cite{PhysRevLett.102.046402,Lin_2011,Chin_2010,Lohse_2015,Nakajima_2016,Song_2019,Ghatak_2020}. Thus the cold atom system may provide an alternative platform to realize our proposal experimentally~\cite{supp}. At last, the competing orders in high-T$_c$ cuprate materials may be rather complex.
Also, various density wave orders are verified to exist in many unconventional superconductors. The relation between the superconductivity and the density-wave order has been an important issue and may relate to the mechanism of the unconventional superconductivity. Moreover,
the interplay between the density-wave orders and the nontrivial topology is of interest and importance, attracting broad interest recently
~\cite{Lin_2021,Hu_2017,Gooth_2019,Wieder_2020,Shi_2021,Qin_2020,Tang_2019}. We expect that our present studies can be extended to other possible competing density-wave orders and other families of unconventional superconductors, providing more insight about the relationship between the density-wave orders, the unconventional superconductivity, and the nontrivial topology.

\section{Summary}

In summary, we have proposed that the higher order topology can be realized in the pseudogap regime of a high-T$_c$ superconducting material coupled with a Chern insulator. With a $d$-wave form pseudogap order ($d$-density-wave order), the gapless edge states in the Chern insulator will be fully gapped. The $d$-density-wave order term acts as an effective mass of the system.
The $d$-wave factor changes sign at the system corner, leading to the higher order topological state.
The higher order topology is confirmed through calculating the edge polarization and the quadrupole moment.
In the underdoped high-T$_c$ superconducting material, the $d$-wave superconducting order may coexist with a $d$-wave density wave order. In this case, the gapless corner states still exist, while each corner contains two zero modes, due to the particle-hole symmetry.

\begin{acknowledgments}
	This work was supported by the NSFC (Grant No. 12074130), the Natural Science Foundation of Guangdong Province (Grant No. 2021A1515012340), and Science and Technology Program of Guangzhou (Grant No. 202102080434).
\end{acknowledgments}

%

\renewcommand{\thesection}{S-\arabic{section}}
\setcounter{section}{0}  
\renewcommand{\theequation}{S\arabic{equation}}
\setcounter{equation}{0}  
\renewcommand{\thefigure}{S\arabic{figure}}
\setcounter{figure}{0}  
\renewcommand{\thetable}{S\Roman{table}}
\setcounter{table}{0}  
\onecolumngrid \flushbottom 

\newpage
\begin{center}\large \textbf{Supplemental material for higher order topological state induced by $d$-wave competing orders in high-T$_c$ superconductor based heterostructure} \end{center}

\section{Experimental realization of the higher order topological state with the high-T$_c$ superconductor platform}
In the main text, we proposed that the higher order topological state may be realized through adding an additional DDW term to the Hamiltonian of the Chern insulator. Now let us discuss in more detail the experimental realization of this proposal. Let us first focus on the heterostructure system.
The schematic illustration is presented in Fig.~S1, namely, a Chern insulator is grown on an underdoped high-T$_c$ cuprate superconductor. Experimentally, the Chern insulator can be realized in the Hg$_{1-x}$Mn$_x$Te/Cd$_{1-x}$Mn$_x$Te quantum well~\cite{Qi_2006} or the intrinsic magnetic topological insulators (Bi,Sb)$_2$Te$_3$~\cite{sci1234414} and MnBi$_{8}$Te$_{13}$~\cite{Hu_2020}.
The DDW order may exist in an underdoped cuprate superconductor~\cite{PhysRevB.62.4880,PhysRevB.63.094503,Forgan_2015}. We expect that in this
heterostructure, the DDW term can be proximity induced to the Chern insulator layer. Then this setup is promising for our present proposal. In the main text, we have shown that the existence of the zero modes can be explored through the LDOS [Fig.~1(c) in the main text]. Experimentally, the LDOS spectra can be measured through the scanning tunneling microscope (STM)
experiments, as illustrated in Fig.~S1.

\renewcommand \thefigure {S\arabic{figure}}

\begin{figure*}[h]
	\includegraphics[width=12.5cm,height=7.5cm]{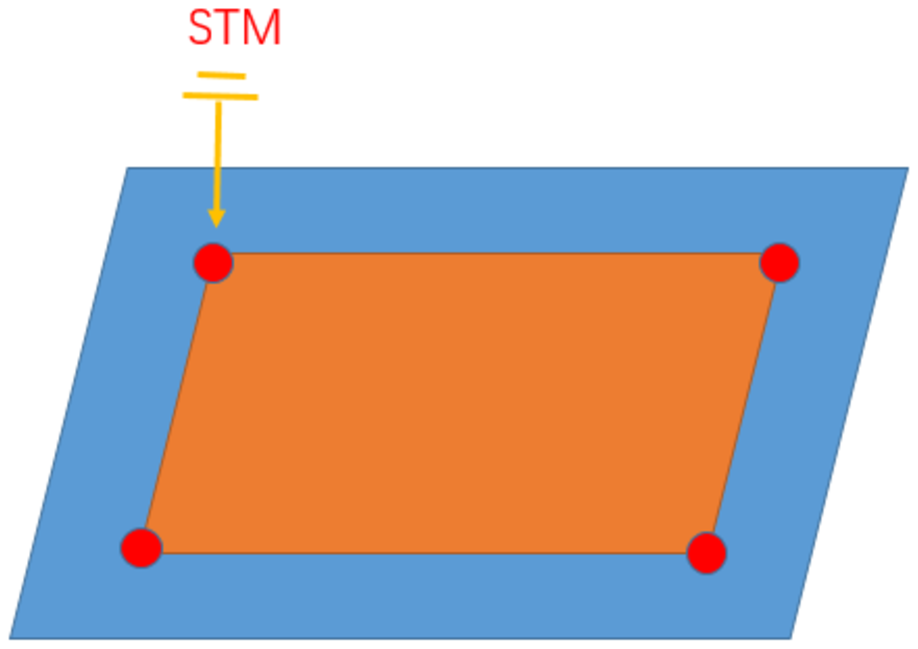}
	\caption{Illustration of the heterostructure. The blue layer represents an underdoped High-T$_c$ superconductor. The brown layer represents a monolayer Chern insulator. The red dots indicate the corner modes.}
\end{figure*}

\section{Experimental realization with the cold atom platform}

As we have mentioned in the main text, the cold atom system is expected to be an alternative platform to realize our proposal.
The spin-orbital coupling has been proposed theoretically and realized experimentally in the cold atom system~\cite{PhysRevLett.102.046402,Lin_2011}. The $s$-wave superfluid pairing state is also realized and controlled through the $s$-wave Feshbach resonance~\cite{Chin_2010}. Moreover, it was proposed that both the quantum anomalous Hall effect and the chiral topological
superfluid phase with the Chern number $C=1$ can be realized in the cold atom system within the experimentally accessible parameter regimes~\cite{PhysRevLett.112.086401}.

In the cold atom system, the hopping term can be driven and controlled by the Raman field~\cite{PhysRevLett.107.255301}.
Considering a spatial dependent field $V(x)$ with $V(x)=V_{0}\cos[{\bf Q_x}\cdot R_x]$,
a periodic modulation pattern along the $x$-direction will be generated by such a field.
Using two spatial dependent Raman fields along the $x$-direction and $y$-direction, the effective hopping term with $M_{x}=iM_{0}(-1)^{R_{x}+R_{y}}c^{\dagger}(R_{x},R_{y})c(R_{x}+1,R_{y})+h.c.$, and $M_{y}=-iM_{0}(-1)^{R_{x}+R_{y}}c^{\dagger}(R_{x},R_{y})c(R_{x},R_{y}+1)+h.c.$ can be created artificially.
The additional factor $(-1)^{R_{x}+R_{y}}$ is due to the periodic modulation of the field $V({\bf r})$.
Note that $M_{x}+M_{y}$ is just the DDW term presented in Eq.(11) of the main text. As illustrated in Fig.~S2, when adding this additional DDW term to the chiral topological state proposed in Ref.~\cite{PhysRevLett.112.086401}, the system will become a second order topological one. Moreover, here the $C=1$ topological superfluid state can be realized~\cite{PhysRevLett.112.086401}, therefore, we expect that with the cold atom platform, the stable Majorana zero modes may exist at the system corners.

\begin{figure*}[h]
	\includegraphics[width=14.5cm,height=7.5cm]{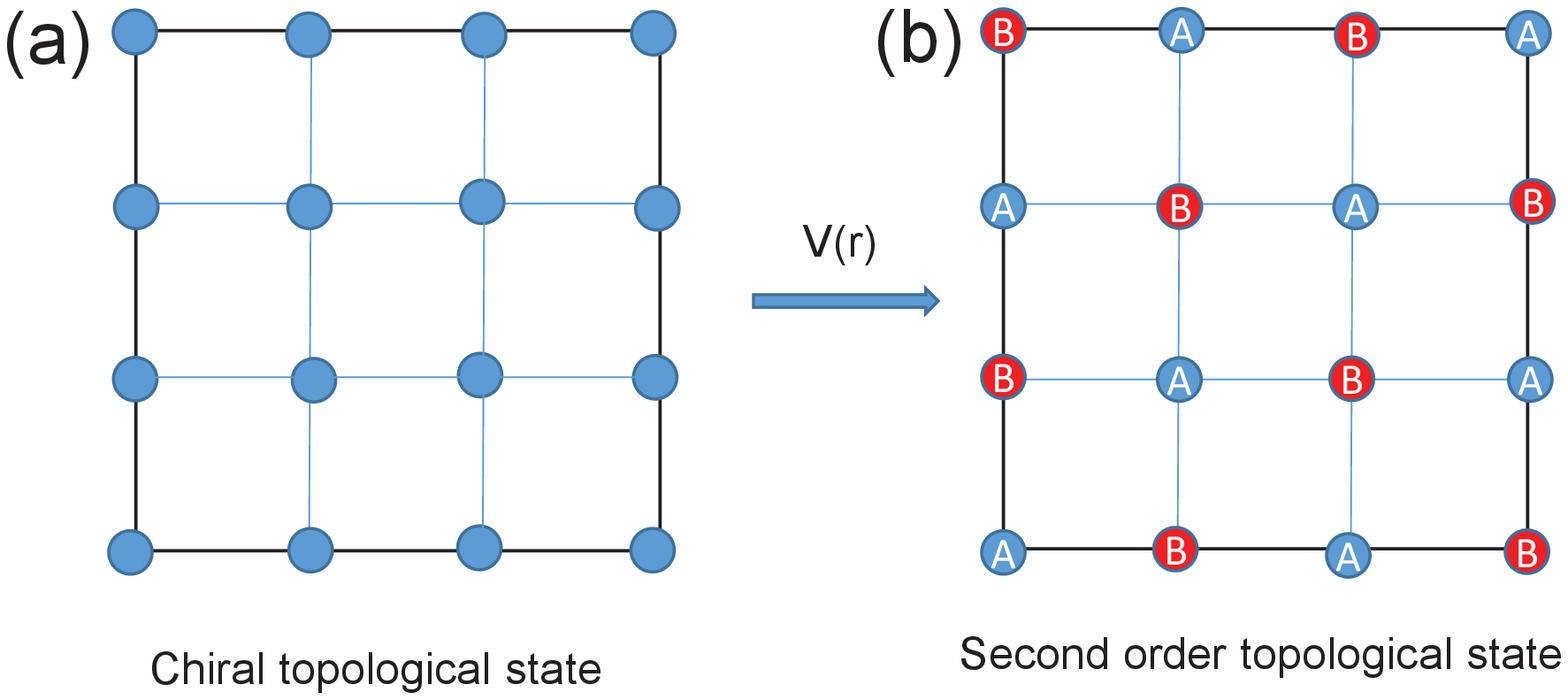}
	\caption{Illustration of the realization of the second order topological state in the cold atom system. The experimental setup for the chiral topological state is proposed in Ref.~\cite{PhysRevLett.112.086401}.}.
	\label{fig:5}
\end{figure*}

\section{Corner charge}
\begin{figure*}[h]
	\includegraphics[width=12.5cm,height=7.5cm]{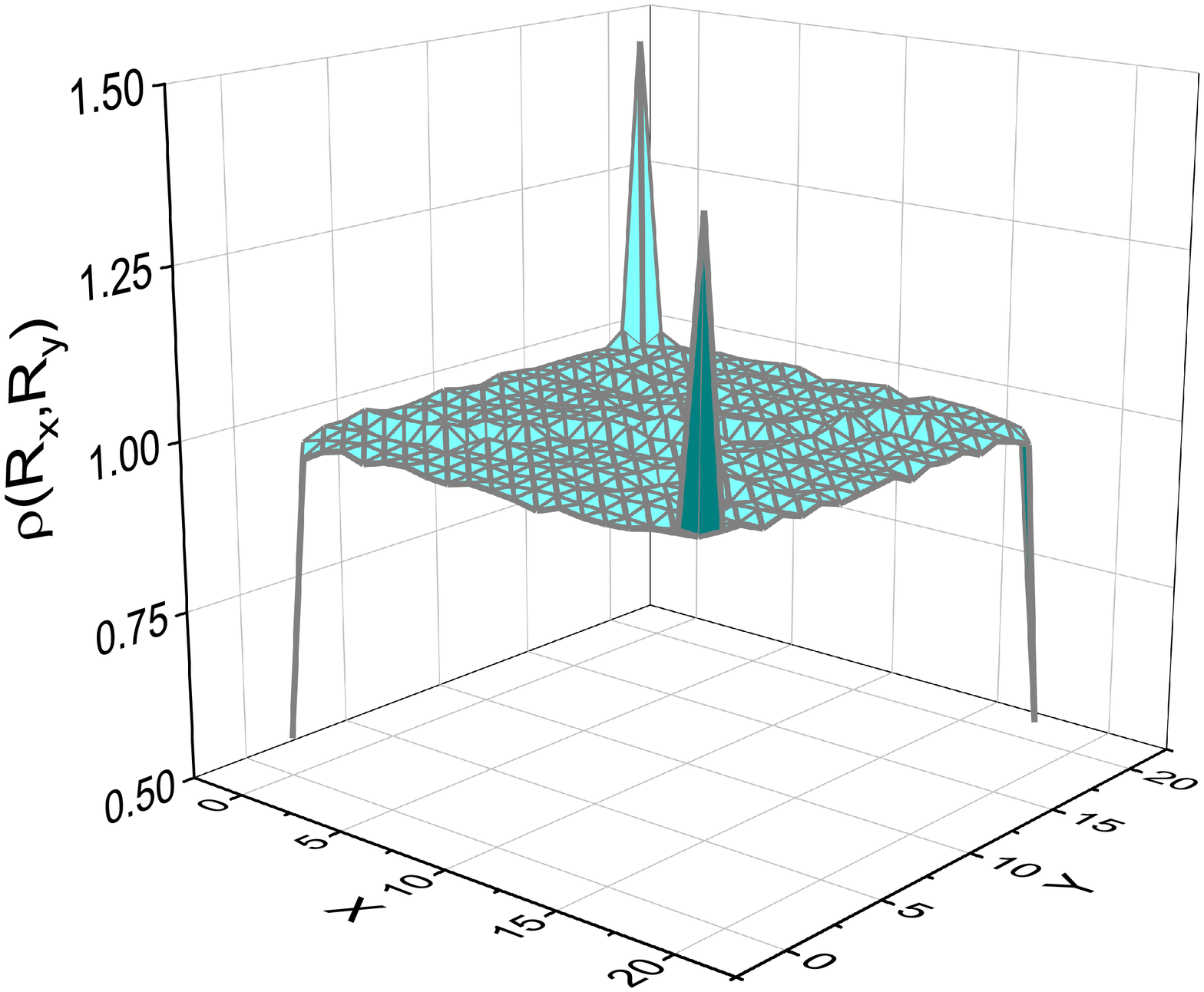}
	\caption{The site-dependent charge density considering the open boundary condition, with the system Hamiltonian being described by Eq.(11) of the main text. The parameters are set as $t=1$, $\lambda_0=1$, $m=1$, and $\Delta_{0}=0.4$. }
	\label{fig:5}
\end{figure*}

For a topologically nontrivial material, it can host fractional charge at the system corner due to the symmetry constrain. The site-dependent charge density is expressed as (using the electronic charge $e$ as the unit),
\begin{equation}
\rho\left(R_x, R_y\right) \equiv  \sum_{n=1}^{N_{\text {occ }}} \left|u^n\left(R_x, R_y\right)\right|^2,
\end{equation}
where $u^n\left(R_x, R_y\right)$ is the n-th component of the eigenstate $\left| u^{n}\right\rangle $ for the Hamiltonian [Eq.(11) in the main text], with only occupied energy bands being considered. The numerical results for the charge density is presented in Fig.~S3. As is shown, the charge density at the system bulk background is 1.0. The corner charge is calculated from the charge density relative to the background charge density. Here the fractional corner charge with  $Q^{\text {corner}}=0.5$ is obtained~\cite{re9,re32}.
The quadrupole moment, the edge polarization and the corner charge should satisfy the following relation~\cite{re9,re32},
\begin{eqnarray}
q_{xy}= p^{edge,y}_{x}+p^{edge,x}_{y}-Q^{\text {corner}}.
\end{eqnarray}
In the main text, we have obtained that $p^{edge,y}_{x}=p^{edge,x}_{y}=0.5$. Then we have $q_{xy}=0.5$, consistent with the result calculated by Eq.(12) of the main text.

\end{document}